\documentclass[12pt]{article}

\usepackage{graphicx}
\usepackage{epsfig}
\usepackage{amssymb,amsmath}
\usepackage{epstopdf}

\newcommand{\be}{\begin{equation}}
\newcommand{\ee}{\end{equation}}
\newcommand{\ba}{\begin{eqnarray}}
\newcommand{\ea}{\end{eqnarray}}
\newcommand{\ii}{{\rm i}}

\newcommand{\bfp}{{\bf p}}
\newcommand{\psib}{\bar{\psi}}

\newcommand{\Oa}{\ensuremath{\mathcal{O}(a)}}
\newcommand{\Ogs}{\ensuremath{\mathcal{O}(g^2)}}
\newcommand{\Oas}{\ensuremath{\mathcal{O}(a^2)}}

\newcommand{\csw}{\ensuremath{{c_\text{\tiny SW}}}}

\newcommand{\eq}{Eq.~}
\newcommand{\eqs}{Eqs.~}
\newcommand{\fig}{Fig.~}

\def\lsi{\raise0.3ex\hbox{$<$\kern-0.75em\raise-1.1ex\hbox{$\sim$}}}
\def\gsi{\raise0.3ex\hbox{$>$\kern-0.75em\raise-1.1ex\hbox{$\sim$}}}
\newcommand{\lsim}{\mathop{\lsi}}
\newcommand{\gsim}{\mathop{\gsi}}

\begin{document}

\title{Cutoff effects of Wilson fermions on the QCD equation of state to \Ogs}

\author{Owe Philipsen and Lars Zeidlewicz\\
Institut f\"ur Theoretische Physik,\\ Westf\"alische Wilhelms-Universit\"at M\"unster,\\ 
48149 M\"unster, Germany}

\date{}

\maketitle
 
\begin{abstract}
We compute the \Ogs{} contribution to the thermodynamic pressure for Wilson fermions 
in the standard, the twisted mass, and clover improved formulation in 
lattice perturbation theory, including finite mass effects.
We compare the continuum approaches of these discretizations for the massive ideal and
interacting gas.
In all cases, for $N_\tau\geq 8$ cutoff effects of Wilson type 
fermions are comparable to those of staggered fermions, but asymptotic scaling requires $N_\tau>10$.
\end{abstract}
 


\section{Introduction}

The most prominent systematic errors in numerical studies of lattice QCD 
are due to cutoff effects. The goal of improvement is to remove the leading lattice corrections for specified observables in a series expansion in the lattice spacing $a$, either by lattice perturbation theory or by nonperturbative means \cite{sym,intro}. The twisted mass fermion formulation represents 
a special case, being automatically $\Oa${} improved at maximal 
twist~\cite{Frezzotti:2003ni}. Corresponding nonperturbative quenched studies are reported in~\cite{Jansen:2005kk}.

For studies of the QCD equation of state at finite temperatures, 
improvement is particularly relevant because a large fraction of the thermally 
distributed field modes lives on the scale $\sim T=1/{aN_\tau}$,
i.e.~on the scale of the cutoff.  In perturbation theory, the pressure can be expressed as $p = p^{(0)}_G+p^{(2)}_G+\ldots + p^{(0)}_F+p^{(2)}_F+\ldots$ where $p_G^{(n)}$ and $p_F^{(n)}$ denote the pure gauge and fermionic contributions to the order of the coupling $g^n$, respectively. Despite existing three-loop results for the free energy at zero temperature~\cite{Athenodorou:2007hi}, perturbative studies of Wilson-type fermions at finite temperature are restricted to 
the ideal gas and chiral limits $p_F^{(0)}(m=0)$; cf.~\cite{hegde} and references therein. 
However, these limits  are far from the phenomenologically
relevant situation,
$T \lsim 4T_c$, and with quark masses still heavier than physical 
in many cases~(cf.~\cite{DeTar:2008qi} for a recent review). 
For twisted mass fermions
there is an investigation only at vanishing temperature and at tree level~\cite{Cichy:2008gk}.

In this paper, we evaluate the effects of nonzero quark masses and interactions to 
order \Ogs{} on the size of cutoff effects in the thermodynamic pressure
by means of lattice perturbation theory. It is well known that the leading cutoff
effects of interacting staggered fermions to the pressure are of \Oas, whereas
massive, interacting Wilson fermions in general will have \Oa{} effects. 
Clover improved Wilson fermions and twisted mass fermions both lead to an \Oa{} improvement 
if the respective parameters are tuned to a suitable value. 
Here, we explicitly evaluate the next-to-leading order (NLO) contribution to the pressure
for these discretizations, and quantitatively compare their cutoff effects with previous
results for staggered fermions~\cite{Heller:1999xz}. 
We find that the interactions increase cutoff effects and thus should not be neglected for making an optimal choice of improvement scheme.
To leading order in the interaction and for lattices with $N_\tau\geq 8$, 
Wilson fermions are competitive with staggered fermions.  

\section{Wilson fermion actions \label{act}}

The twisted mass formulation is obtained from the standard Wilson action $S_W$ for $N_f=2$ degenerate fermions as
\be
S_\text{tm}=S_W(m)+\ii\mu \sum_{x,f,f'}a^4\,\psib_f(x)\gamma_5\tau^3_{ff'}\psi_{f'}(x),
\ee
where we will always work with the standard choice of the Wilson parameter, $r=1$.
$\mu$ denotes the twisted mass parameter
and the diagonal Pauli matrix $\tau^3$ acts in flavour space. 
The bare quark mass in this formulation is $m_q=\sqrt{m^2+\mu^2}$,
\be
 m=m_q\cos(\omega), \quad \mu=m_q\sin(\omega).
\label{massdef}
\ee
Twisted mass fermions provide an automatic \Oa{} improvement over Wilson fermions 
at maximal twist, $\omega=\pi/2$ (see~\cite{Shindler:2007vp}). 
The free quark propagator for twisted mass fermions is 
\be
 \Delta_\text{tm}(p) = a\frac{-\ii\sum_\nu \gamma_\nu \overline{p}_\nu 
+ \frac{1}{2}\hat{p}^2 + am - \ii a\mu\gamma_5\tau^3}{\overline{p}^2
+\left(\frac{1}{2}\hat{p}^2 + am\right)^2 + (a\mu)^2}, 
\label{eq:propagator}
\ee 
with the dimensionless lattice momenta $\overline{p}_\mu = \sin(ap_\mu)$ and $\hat{p}_\mu      = 2\sin(ap_\mu/2)$. 

To identify cutoff effects, the dispersion relation $E(\bfp)$, which is obtained from the poles of the propagator, can be expanded in small lattice spacing,
\begin{equation}
E(\bfp=0) = m_q-\frac{1}{2}am_q^2\cos(\omega) + \Oas.
\label{disp}
\end{equation}
Cutoff effects of \Oa{} set in with quark mass, and their removal at maximal twist  is apparent. 

An improvement scheme that is frequently used in numerical simulations, at both zero and finite temperatures, are clover improved Wilson fermions \cite{sw},
\be
S_\text{SW} = S_W(m)+\csw\; \ii ga^4\sum_{x,\mu,\nu}\frac{r}{4a}
\overline\psi(x)\sigma_{\mu\nu}F^\text{SW}_{\mu\nu}(x)\psi(x),
\ee
with the gluon field strength $F^\text{SW}_{\mu\nu}$.
In this case one uses the propagator of standard Wilson  
fermions, but with rescaled bare quark mass and gauge coupling \cite{bs,wohlertvertex,Capitani:2002mp},
\begin{align}
m_q^\text{SW}&=(m-m_c)(1+b_m\,am),\label{clovscale}\\
g^2_\text{SW}&=g^2(1+b_g\,am_q).\nonumber
\end{align}
Here $m_c(g^2)$ denotes the additive quark mass shift corresponding
to the chiral limit, which as $b_m(g^2)$ and $b_g(g^2)$ 
can be determined in perturbation theory.
The tree-level dispersion relation 
is \Oa{} improved.

The coefficient \csw{} can also be determined in perturbation theory. 
For our purposes, we need these various coefficients to $\mathcal{O}(g^0)$ ($\csw$, $b_g$) or~\Ogs~\cite{Capitani:2002mp,coeffs,lus}. Since these quantities are determined to improve the dispersion relations entering the propagator, they can be applied to the finite temperature case as well as to zero temperature.

\section{The ideal gas limit \label{ideal}}
The thermodynamic pressure is determined by the logarithm of the partition function as $p = (\beta V)^{-1}\ln Z$, i.e. the sum of all bubble diagrams.
The ideal gas contribution of $N_f=2$ free Wilson or twisted mass fermions is immediately obtained from the logarithm of the inverse propagator, where 
as usual the divergent vacuum pressure has to be subtracted by way of renormalization. 
In order to evaluate the one-loop integrals numerically, we compute the corresponding sums for finite volume explicitly, finding that a smooth extrapolation to the thermodynamic limit can be done.

\begin{figure}[t]
\includegraphics[width=0.3\textwidth,angle=-90]{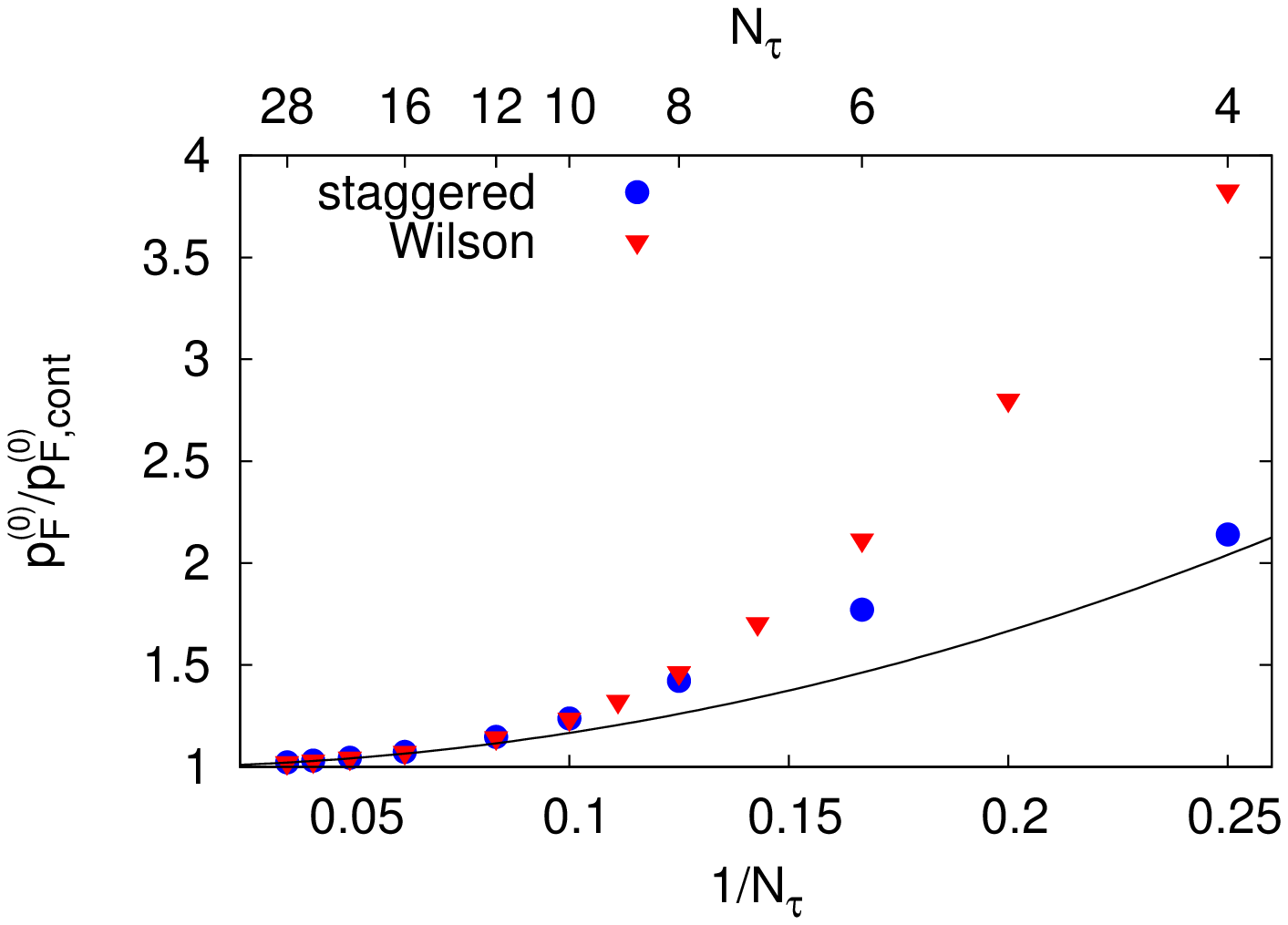}
\hfill\includegraphics[width=0.3\textwidth,angle=-90]{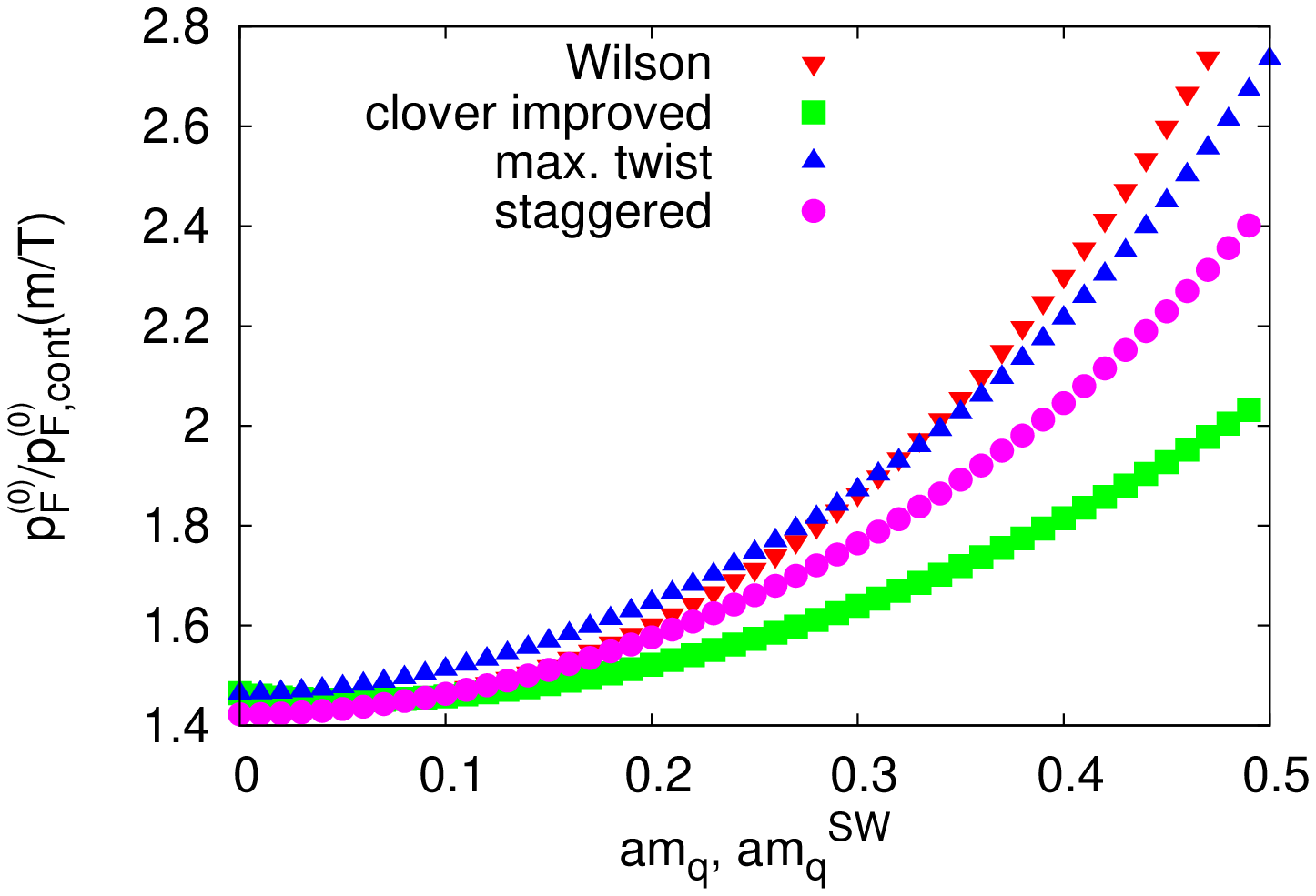}
\caption{\label{free} Left: Stefan-Boltzmann limit for 
massless fermions, normalized to the continuum result. The solid line is the analytic $\mathcal{O}(1/N_\tau^2)$ prediction.
Right: Mass dependence of the Stefan-Boltzmann limit on a lattice with  $N_\tau=8$.
}
\end{figure}
Cutoff effects for the free gas are obtained by comparison with the continuum Stefan-Boltzmann or ideal gas limit, which contains a one-dimensional integral that has to be solved numerically in case of finite mass (cf. also~\cite{Karsch:2000ps}).

In~\cite{hegde} the lattice pressure for quarks has been expanded into a power series in $1/N_\tau$:
\ba
\frac{p}{T^4} &= \frac{(aN_\tau)^3}{2\pi^3}\int\!\!\text{d}^3p\,\ln\left(1+e^{-N_\tau aE(\mathbf{p},m)}\right)
\\
              &\qquad =\, \frac{p^{(a^0)}}{T^4} + \frac{p^{(a^1)}}{T^4}\frac{1}{N_\tau} +  \frac{p^{(a^2)}1}{T^4}\frac{1}{N_\tau^2} + \ldots
\ea
For massless quarks, the coefficients $p^{(a^2)}(m)/T^4$ and $p^{(a^4)}(m)/T^4$  can be calculated in closed form~\cite{hegde} and remarkably are the same for standard staggered and Wilson-type fermions. Differences are introduced only in higher orders of the lattice spacing.
We have repeated this procedure for massive fermions. In this case, one is left with one-dimensional integrals for the expansion coefficients depending on the quark mass that can be solved easily by numerical integration. 

\begin{figure}[t]
\includegraphics[width=0.3\textwidth,angle=-90]{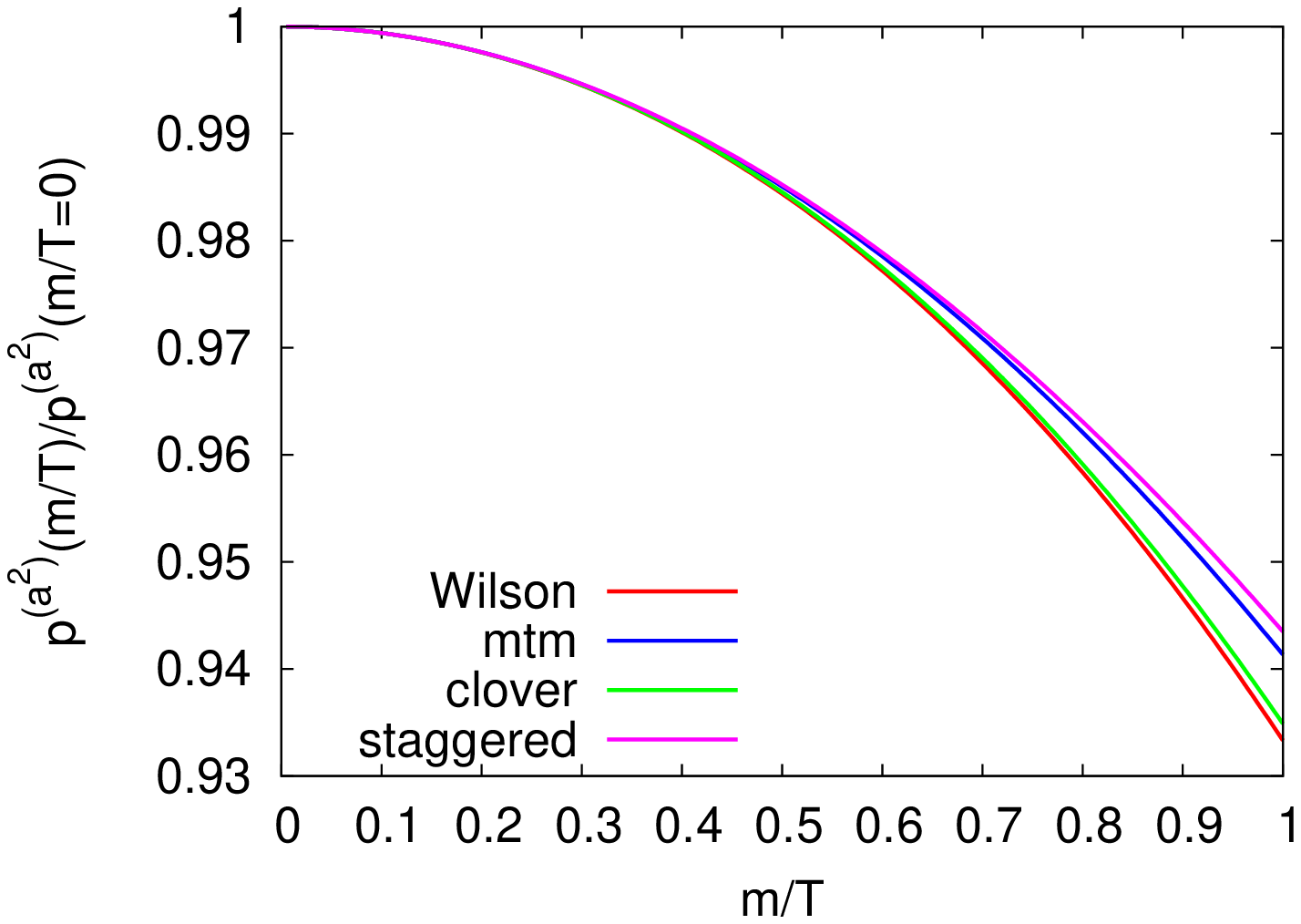}
\hfill \includegraphics[width=0.3\textwidth,angle=-90]{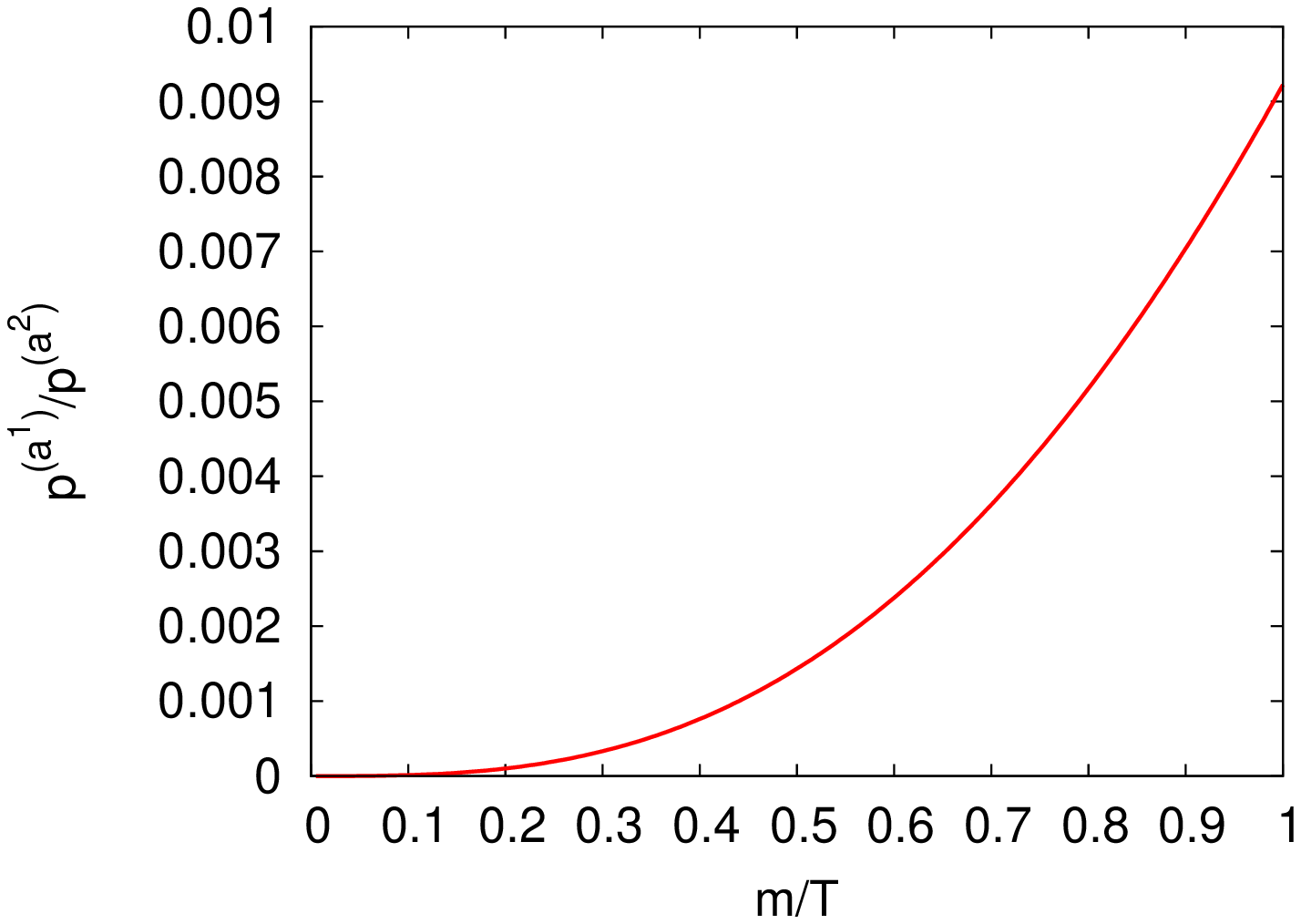}
\caption{\label{freecoeffs} Left: Mass dependence of the \Oas{} contribution to the pressure. Right: Comparison of the first coefficients for the pressure series in $1/N_\tau$.
}
\end{figure}

The fermionic contribution to the pressure on the lattice is shown normalized to the 
continuum contribution
in \fig\ref{free}. 
The chiral limit has been discussed before~\cite{hegde} 
and is reproduced on the left for comparison. 
In this case there is no difference between the 
Wilson formulations; cf.~\eqs(\ref{massdef},\ref{clovscale}). 
Even though in the chiral limit the leading cutoff effects are found to be identical for both types of
discretization,
staggered fermions show a more rapid continuum approach
than Wilson fermions. 
As expected from the dispersion relation, \eq(\ref{disp}), switching on a mass increases 
the cutoff effects for Wilson fermions to \Oa, whereas massive staggered fermions
keep scaling as \Oas{}.
\fig\ref{free} (right) shows that cutoff effects are generally increased by finite quark 
masses.
However, the effect becomes sizeable only for large masses and can be alleviated by 
\Oa{} improvement.

The analytical approach supplements the numerical findings. Up to large quark masses the differences between the \Oas{} contributions for the fermion formulations that have been considered are very small (see Fig.~\ref{freecoeffs}, left). Furthermore, the \Oa{} contribution to pure Wilson fermions is almost negligible when compared to the \Oas{} (see Fig~\ref{freecoeffs}, right). This will only change only once $N_\tau\gsim100$.

\section{Weak coupling expansion to \Ogs}\label{sec:gsq}

The leading \Ogs{} corrections to the fermionic pressure due to interactions are 
given by two diagrams,
\begin{equation}
p_{F}^{(2)}=-\frac{1}{2}\frac{1}{\beta V}\left(
\begin{minipage}{1.3cm}
\includegraphics[width=\textwidth]{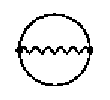}
\end{minipage}
+
\begin{minipage}{2cm}
\includegraphics[width=\textwidth]{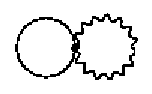}
\end{minipage}\right).
\end{equation}
Twisted mass fermions and ordinary Wilson fermions can be treated simultaneously 
using the propagator \eq(\ref{eq:propagator}), with identical vertices. We use the expressions
as given in \cite{Capitani:2002mp} with the gauge propagator from the standard Wilson plaquette action. 

For purposes of comparison with staggered fermions, we use the corresponding result given in the appendix of \cite{Heller:1999xz}. All integrals have been evaluated numerically using integration routines of the 
\textsc{Cuba}-library~\cite{Hahn:2004fe}. 
Finally, in order to extract the cutoff effects we also need the 
continuum \Ogs{}-corrections from~\cite{Kapusta:1979fh, Kapusta:2006pm}.

Unfortunately, the difference between the finite $N_\tau$ and the vacuum contributions, which is the quantity of interest, shrinks rapidly $\sim 1/N_\tau^4$ and for $N_\tau=8$ is only about 6\,\% of the numerically evaluated integrals, rendering an accurate evaluation difficult.

\begin{figure}[t]
\centerline{
\includegraphics[width=.3\textwidth,angle=-90]{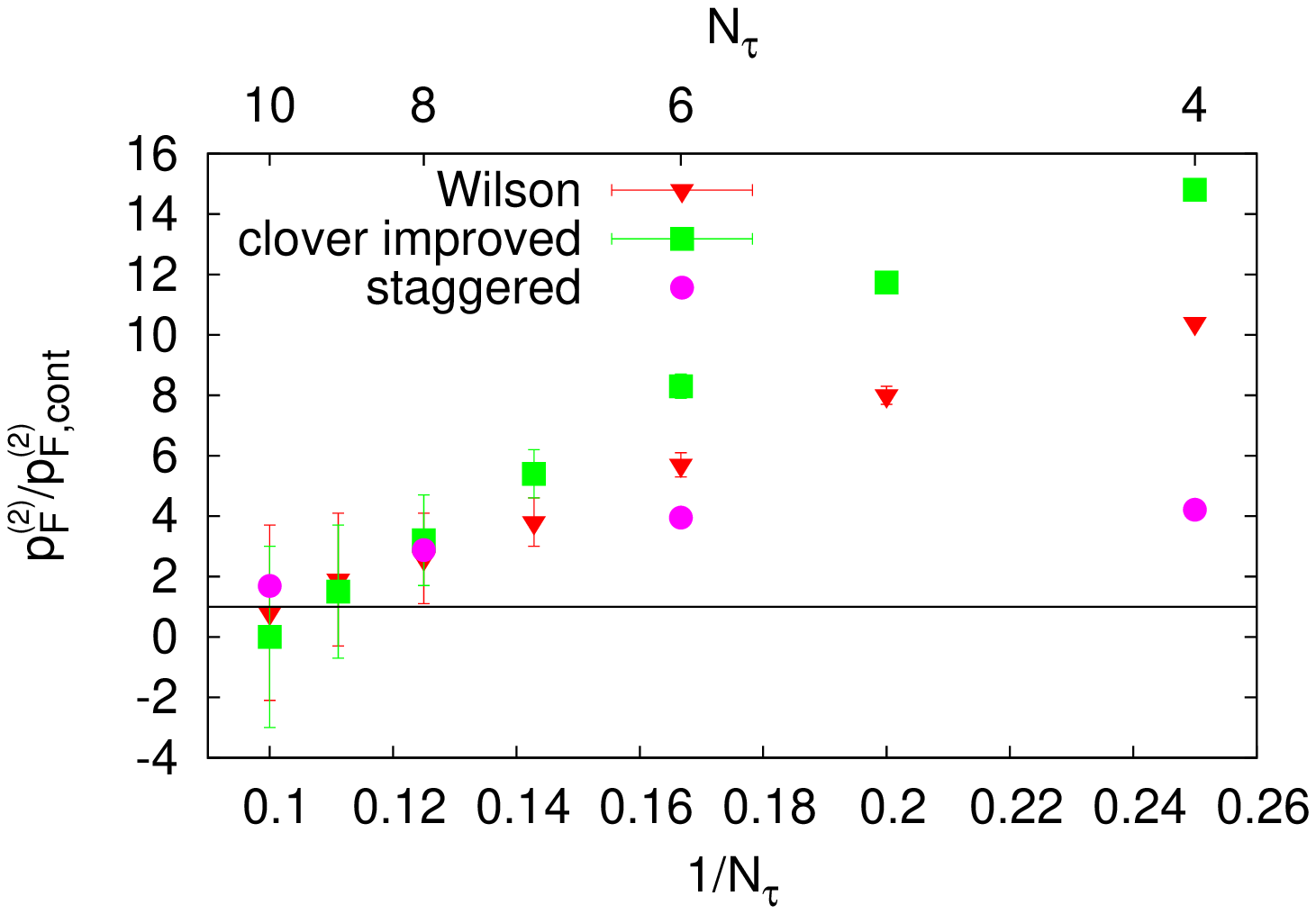}}
\caption{Two-loop contribution to the pressure for massless fermions normalized 
to the continuum result.
}\label{gsq}
\end{figure}

\fig\ref{gsq} shows a comparison of the NLO contribution for Wilson
and staggered fermions in the chiral limit. 
As in the noninteracting case, staggered fermions have smaller cutoff effects
on coarse lattices. 
Comparing the scales of \fig\ref{gsq} and \fig\ref{free} (left), 
interactions indeed appear to make this difference more pronounced. However, 
also for staggered fermions $a^2$ scaling is not setting in until
$N_\tau\geq 8$ at least, by which point the cutoff effects of the 
Wilson discretizations are comparable.

\section{Continuum limit for constant  mass \label{sec:cl}}
For finite mass, mass renormalization leads to an additional contribution to \Ogs{} from the one-loop integrals. For all quark masses, we have $m_R = Z_m m_q$. For Wilson fermions there is additive quark mass renormalization $m_q = m_0-m_c$, while for both the staggered and the twisted masses there is only multiplicative renormalization.
The renormalization constants $Z_m = 1 + Z_m^{(2)} g^2+\ldots$ are in principle discretization dependent. However, the needed first nontrivial coefficient is universal~\cite{Capitani:2002mp}, $Z_m^{(2)} = - 6C_F\ln (a\mu_s)/(4\pi)^2$. Choosing the scale $\mu_s\sim2\pi T$, we find that the scale dependence is small in all cases.

\begin{figure}[t]
\includegraphics[width=.3\textwidth,angle=-90]{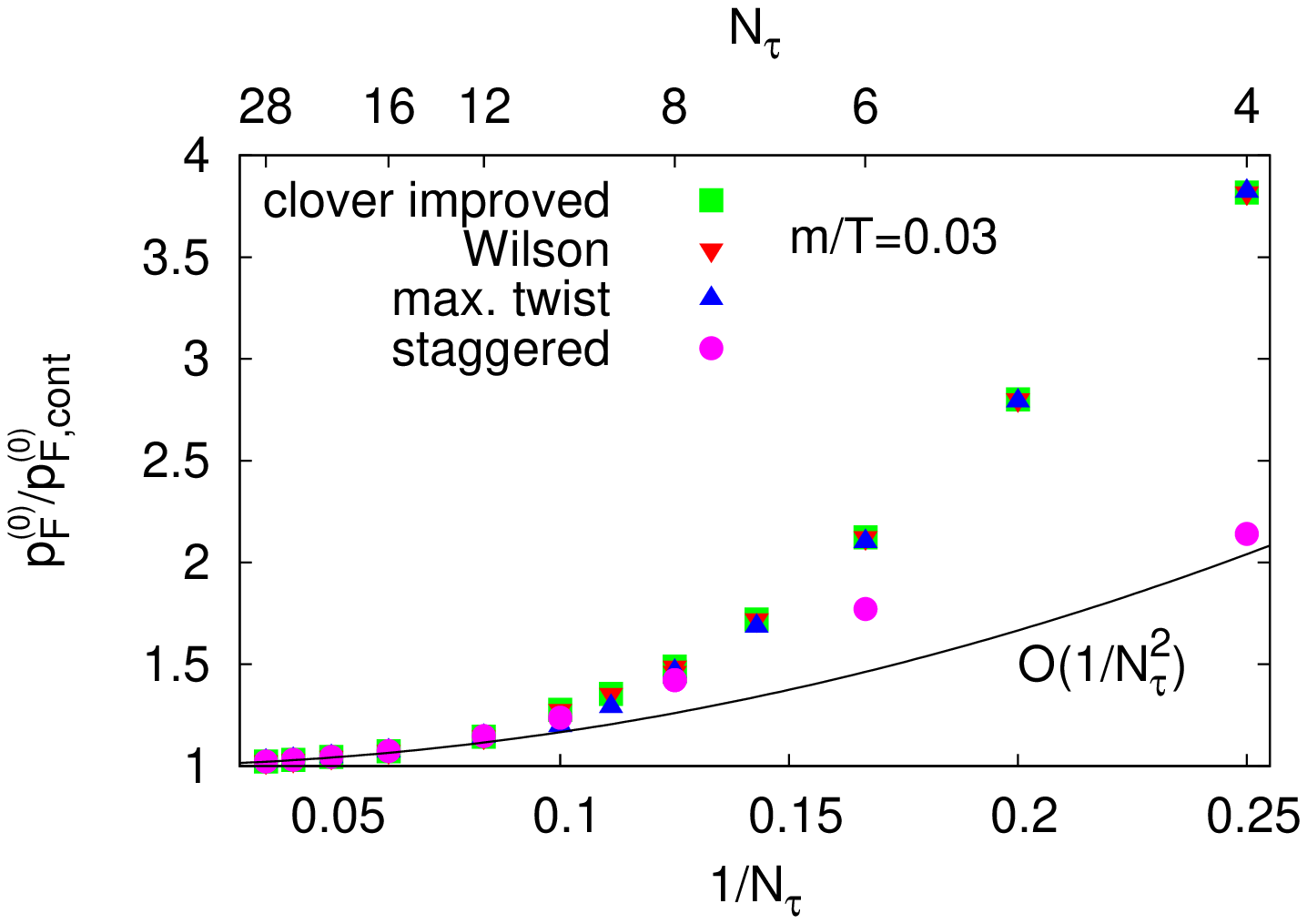}
\hfill\includegraphics[width=.3\textwidth,angle=-90]{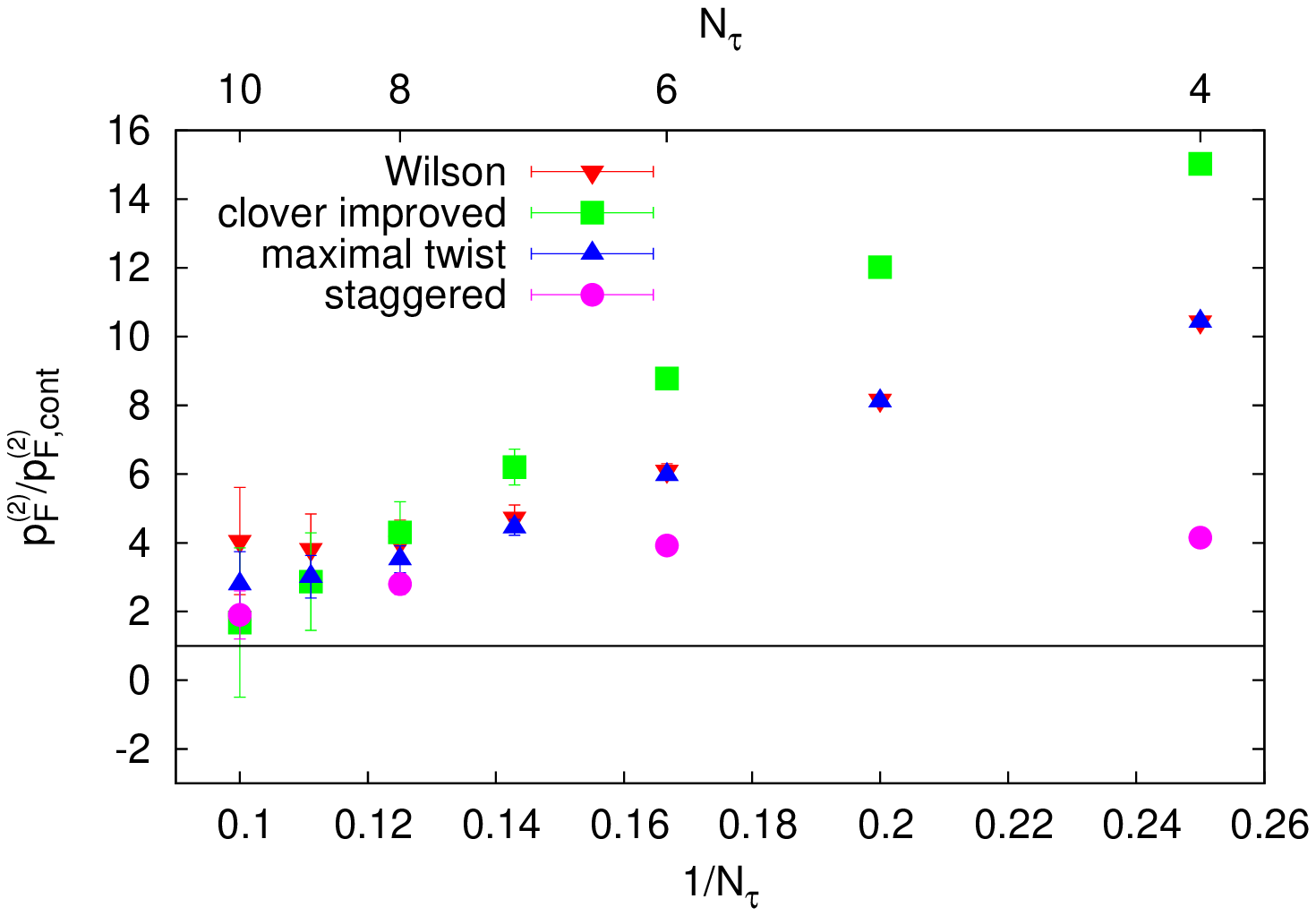}
\caption{Continuum approach of the pressure of massive fermions with a fixed renormalized quark mass $m_R/T=0.03$. 
Left: Ideal gas. Right: Two-loop contribution. \label{pmass}}
\end{figure}

The behavior of massive fermions is shown in Fig.~\ref{pmass}. Despite the rather large numerical uncertainties, one finds a qualitative correspondence to the one-loop case, especially the small dependence on the quark mass and the particular discretization. 
As in the free case, on lattices $N_\tau\geq 8$ Wilson fermions are found to be competitive with staggered fermions. 
However, $a^2$ scaling is not observed for any of the discretizations before $N_\tau\gsim10$.
Up to that, $\mathcal{O}(a)$ effects for unimproved Wilson fermions are negligible.
The comparison of the scales between the free and the interacting situation shows 
that considerations based on the ideal gas limit alone may be insufficient
for the discussion of improvement schemes.
Note that the absolute size of the cutoff effects for the unimproved action appears smaller than for the clover action. This indicates significant higher order contributions for the $N_\tau$ considered here.

\section{Discussion and conclusions}\label{sec:concl}

Our results indicate that the statement made in \cite{hegde} for the massless ideal gas limit,
namely, that an improved dispersion relation leads to the same level of improvement
in the pressure, might generalize to the massive and interacting case.
Let us now discuss the relevance of our results for numerical simulations.
In a regime, where perturbation theory is valid, we can consider the relative cutoff error for the pressure, 
\be
\left|\Delta p_F\right|/p_{F,\text{cont}}=\left|p_{F,\text{lat}}-p_{F,\text{cont}}\right|/p_{F,\text{cont}},
\label{series}
\ee
where $p_{F,\text{cont}}$ and $p_{F,\text{lat}}$ are each calculated to some specified order in the coupling,
in our case \Ogs. In the limit of weak couplings the denominator is dominated by the 
ideal gas limit, and the cutoff effects of the interactions get normalized to that number, rather than 
to $p^{(n)}_{F,\text{cont}}$. In this case they play a rather unimportant role quantitatively, for all discretizations. In fact, the sign of the leading order corrections is negative, so that the relative cutoff
error actually begins to shrink when the coupling is increased from zero.
However, one cannot draw any conclusions from this for the behavior of the pressure near $T_c$.
As the coupling grows, \eq(\ref{series}) quickly develops a pole beyond which results cannot be extrapolated. For longer series the pole might disappear, but the relative cutoff error as a 
rational function will be nonmonotonic in general.

Hence, all one can say in a perturbative analysis is that the lower the temperature, 
the higher the relative importance of  $p^{(n)}_{F,\text{lat}}/p^{(n)}_{F,\text{cont}}$. 
Therefore our analysis suggests that for simulations near $T_c$, Wilson fermions 
scale comparably to standard staggered fermions for fine enough lattices,
$N_\tau\geq 8$.   

We conclude that interactions play a crucial role in discussing improvement schemes
for lattice QCD simulations at finite temperatures. 
An advantage of maximally twisted fermions in this context is that its \Oa{} improvement
appears to hold also nonper\-tur\-batively.
This suggests to further explore the use of this discretization also for finite temperature 
simulations~\cite{Ilgenfritz:2009ns}.
However, independent of the improvement scheme chosen, $a^2$-scaling appears to set in only on lattices $N_\tau>10$.

\section*{Acknowledgements}
We thank S.\,Capitani for several discussions on lattice perturbation theory and
M.~L\"u\-scher for pointing out an error in the treatment of clover fermions in
an earlier version.
This work is supported by Deutsche Forschungsgemeinschaft, grant PH 158/3-1.

\end{document}